\documentclass[aps,twocolumn,prd,epsf]{revtex4}
\usepackage{amssymb}
\usepackage{graphicx}
\newcommand{\be}{\begin{equation}}
\newcommand{\ee}{\end{equation}}
\newcommand{\bea}{\begin{eqnarray}}
\newcommand{\eea}{\end{eqnarray}}
\newcommand{\mpl}{m_{\rm Pl}}
\newcommand{\lpl}{\l_{\rm Pl}}

\begin{document}



\title{An emergent universe from a loop}
\author{David J. Mulryne$^1$, Reza Tavakol$^1$, James E. Lidsey$^1$ and George F. R. Ellis$^2$}

\affiliation{$^1$Astronomy Unit, School of Mathematical Sciences,
Queen Mary, University of London,
London E1 4NS, U.K.}
\affiliation{$^2$Department of Applied Mathematics, University of
Cape Town,Cape Town, South Africa}

\begin{abstract}
\noindent Closed, singularity-free, inflationary cosmological models 
have recently been studied
in the context of general relativity. Despite their appeal, these so called
emergent models suffer from a number of limitations.
These include the fact that they rely
on an initial Einstein static state
to describe the past eternal phase of the universe.
Given the instability of such a state within the context of general relativity,
this amounts to a very severe fine tuning.
Also in order to be able
to study the dynamics of the universe
within the context of general relativity,
they set the initial conditions for the universe
in the classical phase.
Here we study the existence and stability of such models 
in the context of Loop Quantum Cosmology and show that
both these limitations can be partially remedied, once semi-classical
effects are taken into account.
An important consequence of these effects
is to give rise to a static solution (not present in GR),
which dynamically is a centre equilibrium point
and located in the more natural semi-classical regime.
This allows the construction of 
emergent models in which the universe 
oscillates indefinitely
about such an initial static state.
We construct an explicit emergent model of this type,
in which a non-singular past eternal oscillating universe 
enters a phase where the symmetry of
the oscillations is 
broken, leading to an emergent inflationary
epoch, while satisfying all observational and
semi-classical constraints.
We also discuss emergent models in which the universe
possesses both early- and late-time accelerating phases.
\end{abstract}

\maketitle

\section{Introduction}
\setcounter{equation}{0}

One of the fundamental questions of modern cosmology is whether
the universe had a definite origin or whether it is past eternal.
The central paradigm for structure
formation in the universe is the inflationary scenario.
(For a review, see, e.g., Ref. \cite{lidlyth}).
Under very general conditions, inflation is future eternal,
in the sense that once inflation has started, most of the
volume of the universe will remain in an inflating state
\cite{eternal}.
Given certain assumptions, however, it has been argued \cite{borde-etal}
that inflation can not be past null complete -- and
therefore past eternal -- if it is future eternal.

Leading alternatives to inflation that are motivated by
recent advances in string/M--theory are the pre--big bang \cite{pbb,lwc} and
ekpyrotic/cyclic \cite{cyclic} scenarios, respectively.
The fundamental postulate of the
pre--big bang model is that the set of initial data for the universe lies
in the infinite past in the perturbative regime of small string
coupling and spacetime curvature \cite{pbb}. The universe then evolves into
a strongly coupled and highly curved regime before exiting
into the standard, post--big bang phase.

On the other hand, the big bang singularity is interpreted
in the cyclic scenario in terms of the collisions of two
co--dimension one branes propagating in a higher--dimensional spacetime
\cite{cyclic}.
In this picture, the branes undergo an infinite sequence
of oscillations where they move towards and
subsequently away from each other.
From the perspective of a four--dimensional observer,
the collision of the branes is interpreted as the bounce of
a collapsing universe into a decelerating, post--big
bang expansion.

Despite these attractive properties, however, the process that leads to
a non--singular transition between the pre-- and post--big bang phases
is unclear. The search for singularity-free inflationary models
within the context of classical General Relativity (GR) has
recently led to the development of the emergent universe scenario
\cite{Ellis-Maartens,Ellis-Murugun-Tsagas04}. In this model, the universe is
positively curved and initially in a past eternal Einstein static
(ES) state that eventually evolves into a subsequent inflationary phase.
Such models are appealing since they provide specific
examples of non--singular (geodesically complete) inflationary
universes. Furthermore, it has been proposed that entropy
considerations favour the ES state  as
the initial state for our universe \cite{Gibbons87}.

Classical emergent models, however, suffer
from a number of important shortcomings.
In particular, the instability of the ES state
(represented by a saddle equilibrium point in the phase space of the
system) makes it extremely difficult to maintain such a state for an
infinitely long time in the presence of fluctuations,
such as quantum fluctuations, that will inevitably arise.
Moreover, the initial ES state must be set in the classical domain
in order to study the dynamics within the context of GR.
A more natural choice for the initial state
of the universe would, however, be in the semi-classical or
quantum gravity regimes.

A leading framework for a non--perturbative theory
of quantum gravity is loop quantum gravity (LQG).
(For reviews, see, e.g., Ref. \cite{loopreview,loop1}).
Recently, there has been considerable interest in
loop quantum cosmology (LQC),
which is the application of LQG to symmetric states
\cite{bojo}.
Within the framework of LQC, there can exist a `semi--classical' regime,
where spacetime is approximated by a continuous
manifold, but where non--perturbative quantum corrections modify the
form of the classical Einstein field equations \cite{Bojowald2002}.

Motivated by the above discussion, the present paper
studies the existence and nature of static solutions
within the framework of semi--classical LQC in the
presence of a self--interacting scalar field.
Significant progress has recently been made in understanding the
dynamics of this system, and its relevance to
inflation and non-singular behaviour \cite{st, tsm, blmst,
lmnt, mntl, Vereshchagin, Lidsey, Loop_cyclic}.
In addition
to the ES universe of classical cosmology, LQC effects result in a
second equilibrium point in the phase space that we refer to as the
`loop static' (LS) solution. Crucially, this solution corresponds to
a {\em centre} equilibrium point in the
phase space and, consequently, the universe may undergo
a series of (possibly) infinite, non--singular
oscillations about this point. During these oscillations,
the scalar field can be {\em driven up} its potential \cite{lmnt}.
This leads us to consider a new picture for the
origin of the universe, where the universe is initially oscillating
about the LS static solution in the infinite past and eventually
emerges into a classical inflationary era.
The model shares some of the attractive features of the pre--big
bang and cyclic scenarios, in the sense that it is past eternal,
although it exhibits a significant difference in that
the cycles are broken by inflationary expansion. Additionally, unlike
the pre--big
bang and cyclic scenarios,
the emergent model considered here is genuinely non-singular.

The paper is organised as follows. The
existence and nature of static universes in semi-classical LQC
is studied in Section 2. Section 3 discusses the dynamics
that leads to the emergence of an inflationary universe and
a particular inflationary model that is consistent with
present--day cosmological observations is outlined
in Section 4. In Section 5 we discuss a class of
models where the field that drives the pre--inflationary
oscillations may also act as the source of dark energy
in the present--day universe. We conclude with a discussion in Section
6.

\section{Static Universes}

\subsection{Einstein static universe in classical gravity}

Before considering static solutions within the context of LQC,
it is instructive to review the corresponding results for classical
Einstein gravity.

Throughout this paper, we consider an
isotropic and homogeneous Friedmann--Robertson--Walker (FRW)
universe sourced by a
scalar field with energy density and pressure given by
$\rho = \frac{1}{2}\,\dot{\phi}^{2}+V(\phi)$ and
$p = \frac{1}{2}\,\dot{\phi}^{2}-V(\phi)$, respectively,
where $V(\phi )$ represents the self--interaction potential
of the field. The Raychaudhuri equation for such a universe
is given by
\be
\label{Rayclassical}
\frac{\ddot{a}}{a}=-\frac{8\pi \lpl^2}{3}\left[\dot{\phi}^2-V(\phi)\right] \,,
\ee
and local conservation of energy--momentum implies the
scalar field satisfies
\be
\label{KGclassical}
\ddot{\phi} + 3 H \dot{\phi} + V'(\phi) = 0 \, ,
\ee
where a prime denotes differentiation with respect
to the field.
Equations (\ref{Rayclassical}) and (\ref{KGclassical}) together
admit the first integral (the Friedmann equation):
\be
\label{Friedclassical}
H^2=\frac{8\pi \lpl^2}{3}\left [\dot{\phi}^2+V(\phi)\right]
- \frac{K}{a^2}\,,
\ee
where as usual $K=0,\pm 1$ parametrises the spatial
curvature. Combining Eq. (\ref{Friedclassical})
with Eq. (\ref{Rayclassical}) yields
\be
\label{Hdotclassical}
\dot{H}=-4\pi \lpl^2  \dot{\phi}^2  + \frac{K}{a^2}\,.
\ee

The Einstein static universe is characterised by the conditions
$K=1$ and $\dot a = 0 = \ddot a$. In the presence of a scalar field
with a constant potential $V$, it is straightforward to
verify that there exists a unique solution satisfying these
conditions, where the scale factor is given by $a=a_0$ with
$a_0 = (4\pi \lpl^2 {\dot \phi}^2)^{1/2}$.
There are two important points to note regarding this solution.
Firstly, it represents a saddle equilibrium point in
the phase space, which is unstable to linear perturbations.
Secondly, in order to enable the analysis
to be performed within the context of GR,
it is necessary to assume that $a_0$ lies in the
classical domain.

In the emergent universe scenario, it is assumed that
the potential becomes asymptotically flat
in the limit $\phi \to -\infty$ and that the initial conditions
are specified such that the ES configuration
represents the past eternal state of the universe,
out of which the universe slowly evolves into an inflationary phase. However,
the instability of the ES universe
ensures that any perturbations -- no matter how small --
rapidly force the universe away
from the ES state, thereby aborting the scenario.

In the next subsection we shall see that employing
a more general LQC setting can partially resolve both these
issues.

\subsection{Static solutions in semi-classical LQC}

It has recently been shown that in the semi-classical regime
the cosmological evolution equations become modified \cite{Bojowald2002}.
When restricted to FRW backgrounds,
this regime is characterised by the scale factor of the universe
lying in the range $a_i < a< a_*$, where $a_i \equiv \sqrt{\gamma}\lpl$
determines the scale for the discrete quantum nature of spacetime
to become important
and $\gamma \approx 0.274$ is the Barbero--Immirzi parameter \cite{bi},
and $a_*  \equiv \sqrt{\gamma j/3}  \lpl$, where $j$ is
a parameter that arises due to ambiguities in the quantization
procedure \cite{amigpara}. It must take positive, half--integer values
but is otherwise arbitrary \footnote{More specifically,
any irreducible ${\rm SU(2)}$ representation with spin $j$ may be chosen
in the quantization scheme when rewriting the classical scale factor in terms
of holonomies. The fundamental representation corresponds to $j=1/2$.}.
The standard classical cosmology is recovered
above the scale $a_*$ and the parameter $j$ therefore sets the
effective quantum gravity scale.
The dynamics in this regime is determined by replacing the inverse volume
term $a^{-3}$ that arises in the classical matter Hamiltonian
${\cal{H}}_{\phi} = \frac{1}{2} a^{-3} p^2_{\phi} +a^3 V(\phi )$ with
a continuous function $d_j(a)$ that approximates the eigenvalues of the
geometrical density operator in LQC. This `quantum correction' function
is given by $d_{j}(a) \equiv D(q)a^{-3}$, where
\begin{eqnarray}
\label{defD}
D(q) &=& \left(\frac{8}{77}\right)^6 q^{\frac{3}{2}}
\left\{ 7\left[(q+1)^{\frac{11}{4}} - |q-1|^{\frac{11}{4}}\right]
\right. \nonumber \\
&-& \left. 11q\left[(q+1)^{\frac{7}{4}}- {\rm sgn} (q-1)|q-1|^{\frac{7}{4}}
\right]\right\}^6 \,,
\end{eqnarray}
and  $q \equiv (a/a_*)^2$. As the universe evolves through
the semi--classical phase, this function
varies as $D \propto a^{15}$ for $a \ll a_*$, has a global
maximum at $a \approx a_*$, and falls monotonically to $D = 1$
for $a > a_*$.

The effective field equations then follow from the Hamiltonian
\cite{Bojowald2002}:
\begin{equation}
\label{semiclassham}
\hat{\cal{H}} = -\frac{3}{8\pi \ell^2_{\rm Pl}}
\left( \dot{a}^2 +K \right) a
+ \frac{1}{2}
d_{j} p^2_{\phi} +a^3V =0 \,,
\end{equation}
where $p_{\phi} = d^{-1}_{j} \dot{\phi}$ is the momentum
canonically conjugate to the scalar field.
The Raycharduri equation becomes
\be
\label{Rayquantum}
\frac{\ddot a}{a} = - \frac{8 \pi \lpl^2}{3 D} \, \dot \phi^2 \,
\left(1 - \frac{1}{4}\frac{d\ln D}{d\ln a} \right)
+ \frac{8 \pi \lpl^2}{3} \, V(\phi) \,,
\ee
and the modified scalar field equation takes the form
\be
\label{KGquantum}
\ddot{\phi} = - 3 H \left( 1- \frac{1}{3} \frac{d \ln D}{d \ln a} \right)
\dot{\phi} - D V' \, .
\ee
The first integral of Eqs. (\ref{Rayquantum}) and (\ref{KGquantum})
is given by the modified Friedmann equation
\be
\label{Friedquantum}
H^2=\frac{8\pi \lpl^2}{3}\left [\frac{\dot{\phi}^2}{2D}+V(\phi)\right]
- \frac{K}{a^2} \,,
\ee
where in the LQC context $K$ can only take
values $0,+1$. Combining Eq. (\ref{Rayquantum}) with
Eq. (\ref{Friedquantum}) then implies that
\be
\label{Hdotquantum}
\dot{H} =  -\frac{4\pi \lpl^2 \dot{\phi}^2}{D}
\left( 1-\frac{1}{6} \frac{d \ln D}{d \ln a}  \right) +\frac{K}{a^2} \,.
\ee

Equation (\ref{Rayquantum})--(\ref{Hdotquantum})
can be rewritten in the form of a three--dimensional
dynamical system in terms of variables $ \{ a,H,V \}$
\cite{Vereshchagin}. (The present study
extends the results of \cite{Vereshchagin} and, more importantly,
provides an analytical account of the dynamics).
Employing the Friedmann equation (\ref{Friedquantum})
to eliminate the scalar field's kinetic term, and assuming that
$dV/d\phi$ can be expressed as a function of the potential,
allows the complete dynamical system
to be described by equations:
\begin{eqnarray}
\label{2dsystem}
\dot{H}&=&\left ( 8 \pi \lpl^2 V -3H^2 \right )
\left( 1-\frac{A}{6} \right ) \nonumber\,\\
&+& \frac{K}{a^2} \left ( \frac{A}{2}-2\right ) \,,\\
\label{2dsystem1}
\dot{a}&=&Ha\,,\\
\label{Vdot}
\dot{V}&=& V'(V) \left ( \frac{6 D H^2}{8 \pi \lpl^2}
- 2DV + \frac{6 D}{8 \pi \lpl^2 a^2} \right )^{\frac{1}{2}}\,,
\end{eqnarray}
where the function $A (a) \equiv d \ln D/d \ln a$.
In this Section, with the aim of considering
emergent universes we shall take
$K=+1$ and initially
consider constant potentials. In this case,
Eq. (\ref{Vdot}) is trivially satisfied and the system reduces to the
two--dimensional autonomous system (\ref{2dsystem})--(\ref{2dsystem1}).

The static solutions ($\dot a = 0 = \ddot a$) then
correspond to the equilibrium points of this system,
which are given by
\be
\label{eq-points}
H_{\rm eq}=0, \qquad a=a_{\rm eq} \,,
\ee
where $a_{\rm eq}$ is given by solutions to the constraint equation
\be
\label{AB}
A (a_{\rm eq}) = B (a_{\rm eq}) , \qquad
B (a) \equiv  \frac{6(8 \pi \lpl^2
V a^2 - 2)}{(8 \pi \lpl^2 V a^2 - 3)} \,.
\ee
It follows from Eq. (\ref{Hdotquantum}) that the field's kinetic
energy at the equilibrium points is given by
\be
\label{KEfix}
\dot{\phi}_{\rm eq}^2
= \frac{D_{\rm eq}}{4\pi {a_{\rm eq}}^2 \lpl^2 (1-A_{\rm eq}/6)} \,.
\ee

Equation (\ref{AB}) implies that a necessary and sufficient
condition for the existence of the equilibrium points is that
the functions $A(a)$ and $B(a)$ should intersect.

\begin{figure}[!t]
\includegraphics[width=7.2cm]{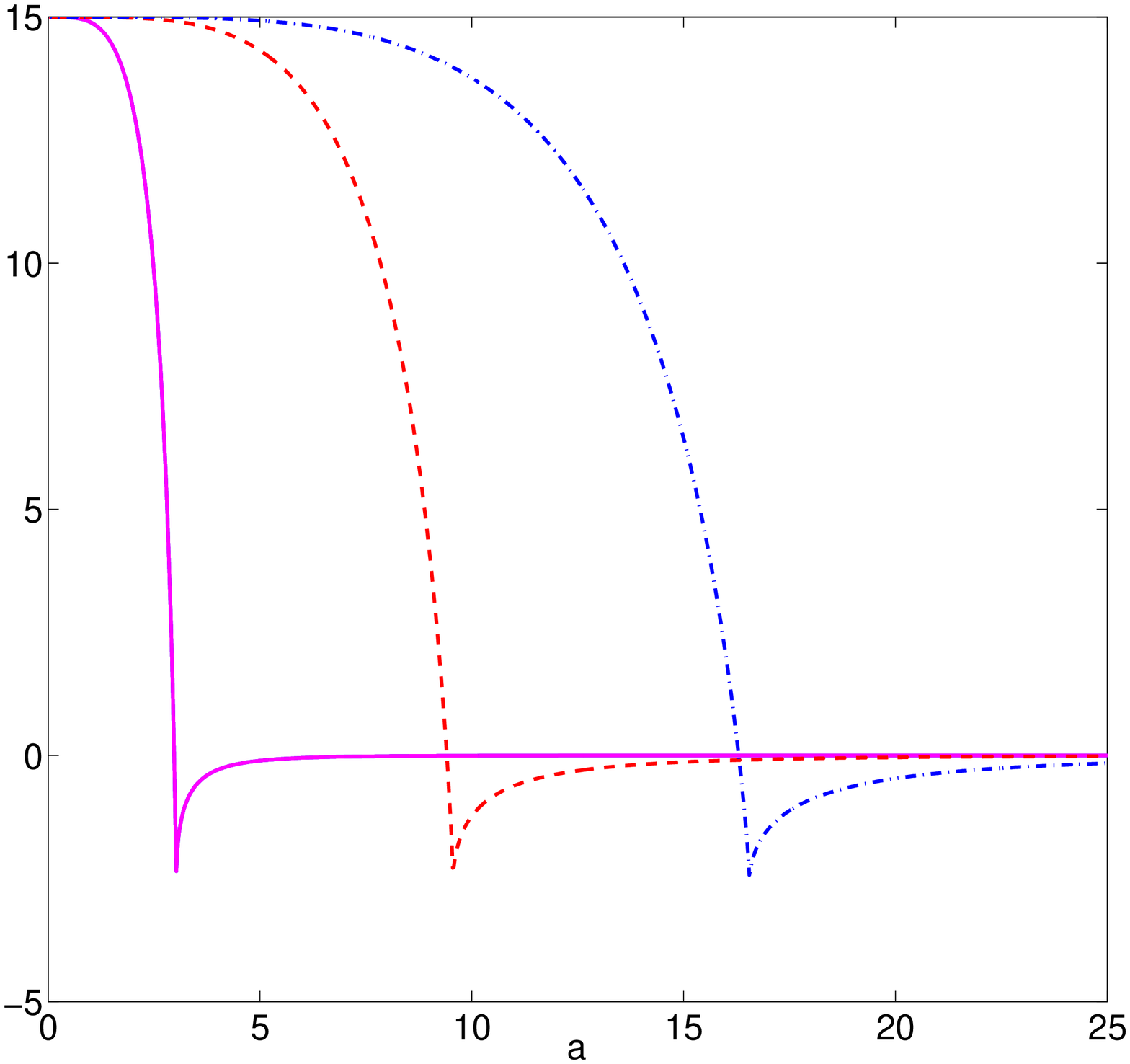}
\includegraphics[width=7.2cm]{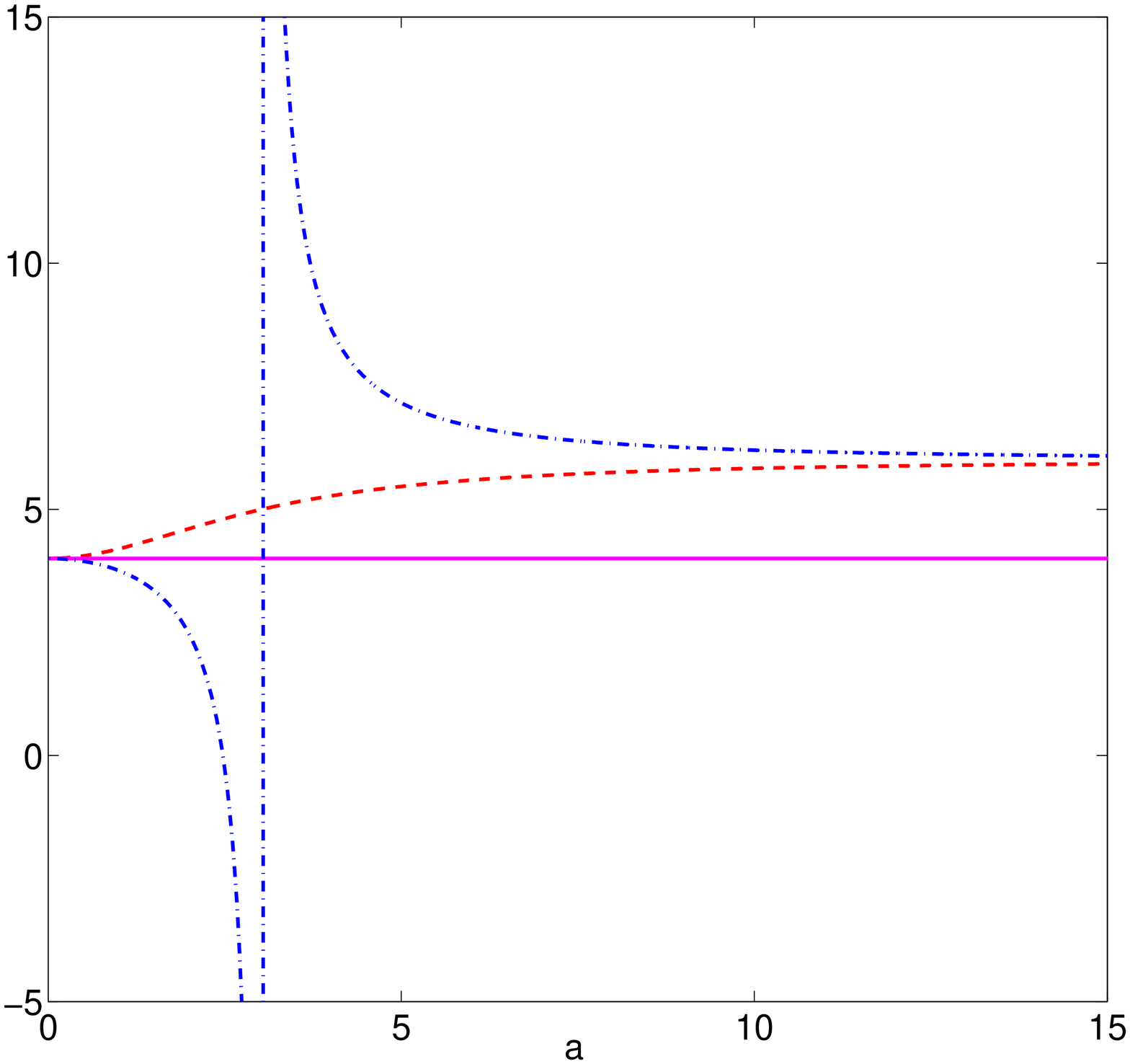}
\caption[] {\label{fig1.eps}
The top panel shows the plots for the function $A$ against $a$ for $j=100$
(solid line), $j=1000$ (dashed line) and $j=3000$ (dot-dashed line).
The bottom panel shows plots of the function $B$ against
$a$ for $V=0$ (solid line), $V>0$ (dot-dashed line) and $V<0$ (dashed line).
Note the dramatic change in the shape of the function as
the sign of $V$ is altered.
}
\end{figure}

Here we are interested in how the properties of the
equilibrium points alter as the
potential is varied. This is best seen
by considering how the functions
$A(a)$ and $B(a)$ (and hence their points of
intersection) change. The function $A$ has the form shown
in the top panel
of Fig.~\ref{fig1.eps}. It asymptotes to the
constant value $A=15$ at $a=0$,
decreases to a minimum value of $A_{\rm min}=-5/2$ at $a=a_*$, and
then asymptotes to zero from below as $a \rightarrow \infty$.
Increasing the parameter
$j$ increases the value of $a_*$ and this results
in moving the turning point of the function $A$ to
larger values of the scale factor.
The important point to note, however,
is that the qualitative form of the function $A$ remains
unaltered for all values of $j$,
as can be seen from Fig.~\ref{fig1.eps}.

The qualitative
nature of the function $B$, on the other hand,
is sensitive to the sign of the
potential, as can be seen from the bottom panel of Fig.~\ref{fig1.eps}.
We briefly discuss the behaviour of $B$ for each case in turn.
For $V=0$, it is given by the (solid) horizontal line $B=4$. For $V>0$ it
represents a hyperbola (dot-dashed curve) with a single
(since the scale factor $a$ is non-negative) vertical asymptote
given by
\be
\label{asymptote}
a= \sqrt{\frac{3}{8 \pi \lpl^2 V}} \,.
\ee
The region to the left of this asymptote
defines the region in the
$\{ A/B,a \}$ plane where the reality condition,
${\dot \phi}^2>0$, is satisfied. Thus,
the upper--right branch of the hyperbola
plays no role in determining the existence of the equilibrium points.
In the limit of $a \rightarrow 0$, the function $B \rightarrow 4$ which
coincides with the value of the function in
the case $V=0$. As $a$ is increased,
the function takes progressively smaller values as
the vertical asymptote is approached.
As $V$ tends to zero from above, it causes
the asymptote to move to progressively larger values of $a$,
and we may therefore formally
view the $V=0$ case as the limit
where the asymptote moves to infinity.
Finally, for negative potentials the qualitative behaviour
of the function $B$ is changed to 
the dashed line in the bottom panel of Fig.~\ref{fig1.eps}
and there is no vertical asymptote. 
As in the case of positive 
potentials, $B \rightarrow 4$ as $a\rightarrow 0$,
but the function $B$ now increases monotonically as
$a$ increases, ultimately tending to the value $6$ as $a \rightarrow
\infty $. As the potential tends to zero from below, the
function $B$ still tends to the asymptotic value $B=6$,
but at larger $a$.
An important point to note is that for {\em all}
values of $V$, the function $B$ satisfies the condition $B<6$
(for physically relevant regions of the $\{ A/B,a \}$ plane),
and hence {\em equilibrium points can only occur when $A=B<6$}.

Once the positions of the equilibrium points have been determined,
their nature can be found by linearising about these points.
The eigenvalues are given by
\be
\lambda^2= \left [\frac{4-A}{a^2}
+ \frac{1}{6 a (1-A/6)}\frac{dA}{da} \right ]_{a_{\rm eq}} \, .
\ee
When $\lambda^2 <0$, this leads to imaginary
eigenvalues and implies that the equilibrium point is a
centre, whereas the point is a saddle when $\lambda^2 >0$.
Although this expression is rather complicated,
it can be verified that for $A<6$ (which is
a necessary condition for the existence of equilibrium points),
$\lambda^2$ is negative for $a<a_*$ and positive for
$a>a_*$. Hence we have the important result that,
{\em equilibrium points occurring in the semi-classical 
regime are centres and those occurring in the 
classical regime are saddles}.

To determine the existence of equilibrium points, we again consider the
different signs of the potential separately.

\begin{figure}[!t]
\includegraphics[width=7.2cm]{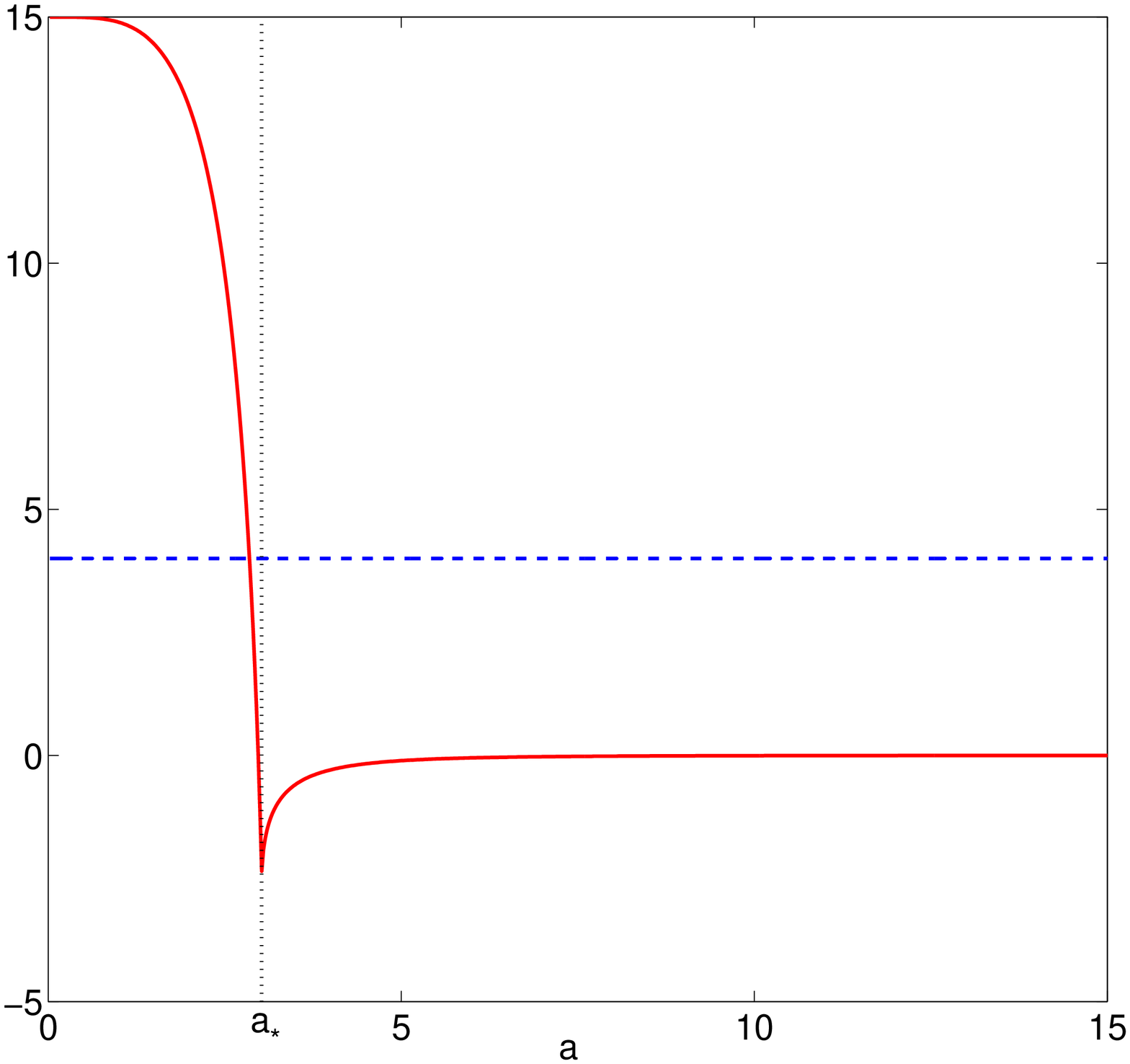}
\includegraphics[width=7.2cm]{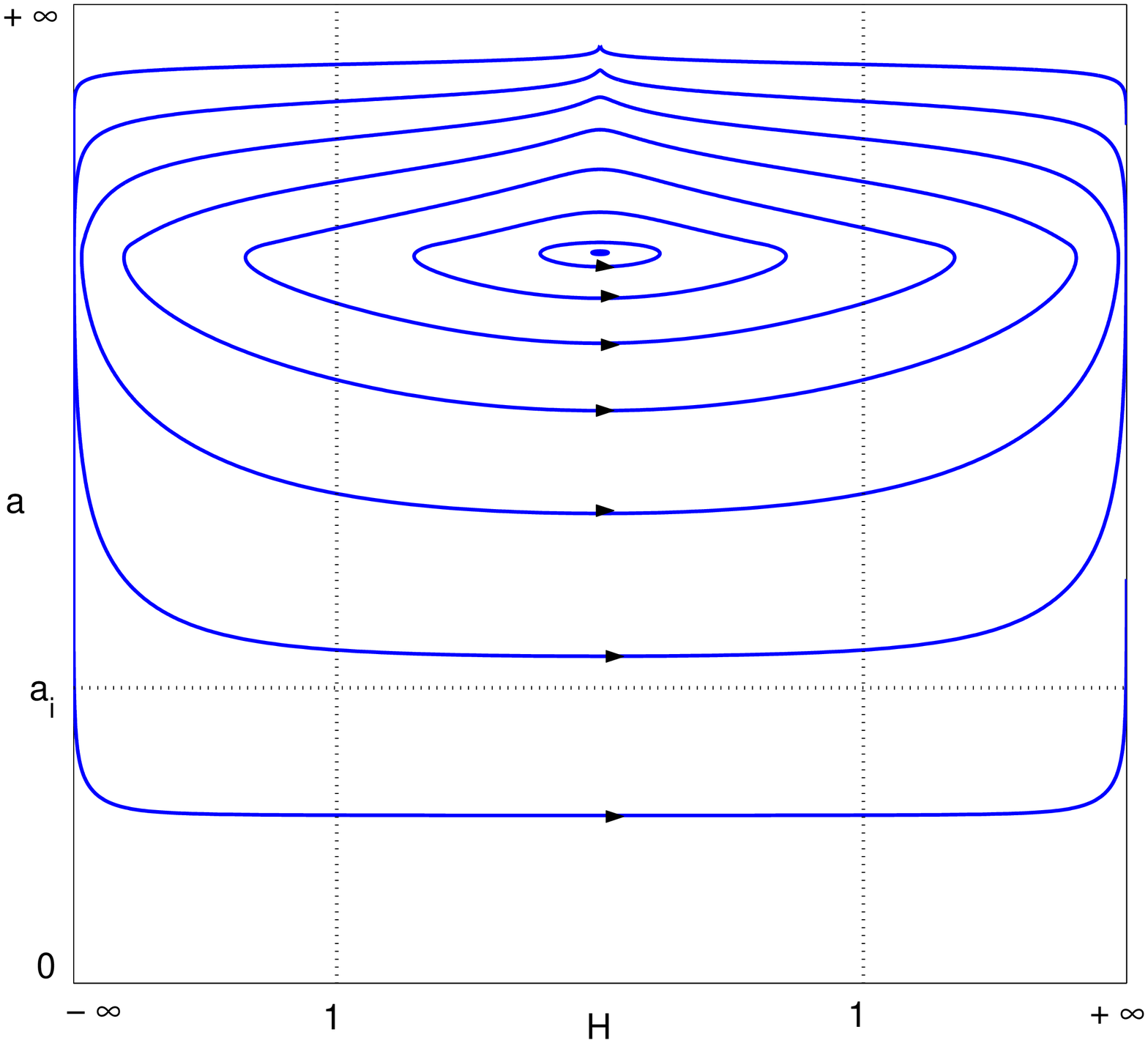}
\caption[] {\label{fig2.eps}
The top panel shows the plots of functions $A$ (solid line) and
$B$ (dashed line) against
$a$ for $V=0$. The vertical dotted line 
denotes the position of $a=a_*$.
The bottom panel is the corresponding phase portrait demonstrating the
centre equilibrium point that occurs in this case.
The plot is compactified using
$x={\rm arctan} (H)$ and $y= {\rm arctan}(\ln a)$, in order
to present the entire phase space.
The vertical dotted lines in the bottom panel demarcate the
region in which the Hubble parameter is
less than the Planck scale.
The horizontal dotted line marks $a_i$, where
the quantum regime begins.
All axes are in
Planck units, and $j=100$.

}
\end{figure}

$V=0$:
In the case of a massless scalar field with $V=0$, the condition (\ref{AB})
has a single solution given by $A=4$, implying a single
equilibrium point.
The position of this equilibrium point can be seen
from the intersection point in the top panel of Fig.~\ref{fig2.eps}.
Since this point occurs at $a<a_*$,
it is a centre, with its phase portrait consisting of a
continuum of concentric orbits (see the bottom panel of Fig.~\ref{fig2.eps}).

\begin{figure}[!t]
\includegraphics[width=7.2cm]{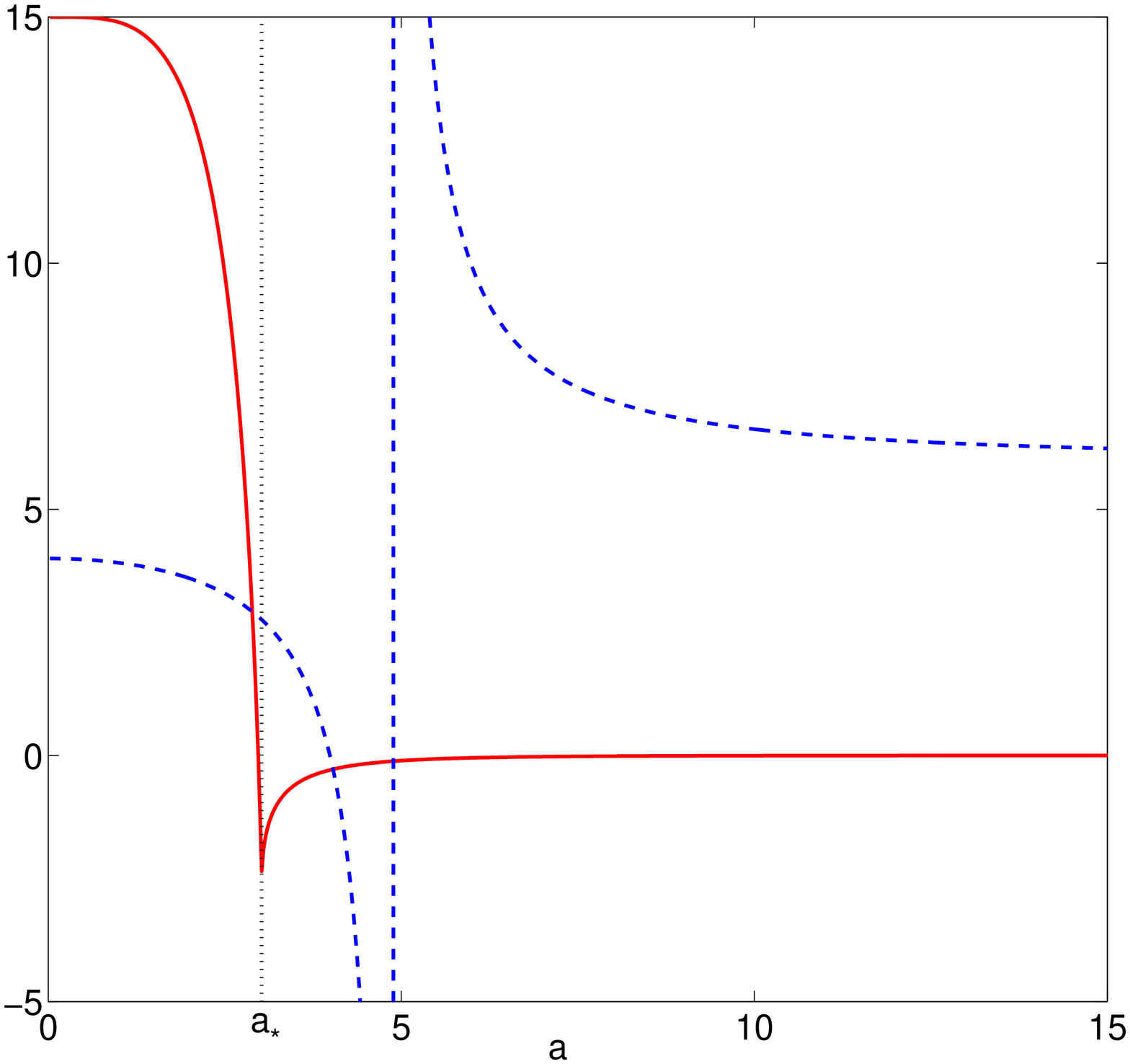}
\includegraphics[width=7.2cm]{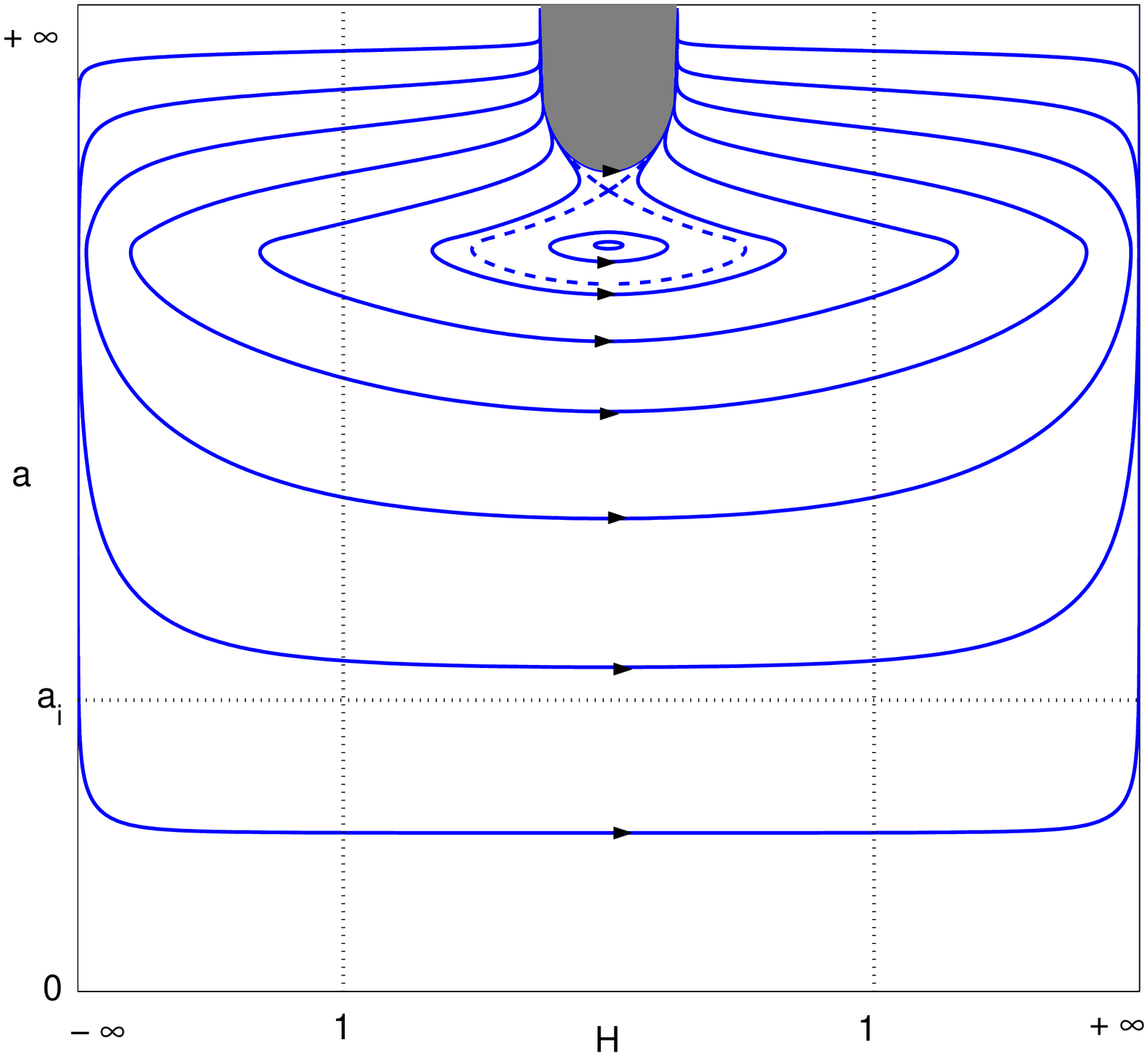}
\caption[] {\label{fig3.eps}
The top panel shows the plots of functions $A$ (solid line) and
$B$ (dashed line) against
$a$ for $V=0.005$.  For this value of $V$,
the equilibrium points are close to coalescing. 
The vertical dotted line denotes the position of $a=a_*$.
The bottom panel is the corresponding phase portrait demonstrating the
centre equilibrium point and saddle point which occurs in this
case with the dashed line indicating the separatrix.
The shaded area
is the unphysical region where $\dot{\phi}^2$ would be negative.
The plot is compactified as in Fig. \ref{fig2.eps}.
The dotted lines in the bottom panel
are as described in Fig. \ref{fig2.eps}.
All axes are in
Planck units, and $j=100$.
}
\end{figure}

$V>0$:
As described above, for small positive values of the potential,
the vertical asymptote of the function $B$ is
far to the right of the origin,
which results in two points of intersections
between the functions $A$ and $B$ (see the top panel of Fig.~\ref{fig3.eps}).
There are therefore {\em two} equilibrium points in this case:
the first occurs at $a<a_*$ and hence is  a centre,
while the second  occurs at $a>a_*$ and is therefore a saddle.
As the potential is increased, the asymptote moves towards the origin
causing the equilibrium points to eventually coalesce
when $V=V_*$. This occurs at the point of
tangency of the functions $A$ and $B$, i.e. at
the minimum (which is a cusp) of
the function $A$ located at $a=a_*$. Thus, 
using $B$ and the fact that
$A|_{a_*}=-5/2$, we obtain
\be
\label{potentialmerge}
V_* = \frac{39}{136 \pi \lpl^2 a_*^2}
=  \frac{117}{136 \pi \gamma j} m_{\rm Pl}^4 \,.
\ee

For values of $V>V_*$,
the asymptote moves further towards the origin. 
Consequently, the functions
$A$ and $B$ will no longer intersect in the part
of the plane which satisfies the reality condition,
and therefore there will be
no equilibrium points in this case.
It is also easy to understand the effect of
changing the parameter $j$ on $V_*$.
Increasing $j$ increases $a_*$, 
which means that the point of tangency occurs for
smaller values of the potential.
This implies that larger values of the quantization parameter, 
and hence $a_*$, lead to smaller
values of the potential beyond which no equilibrium
points can exist.

\begin{figure}[!t]
\includegraphics[width=7.2cm]{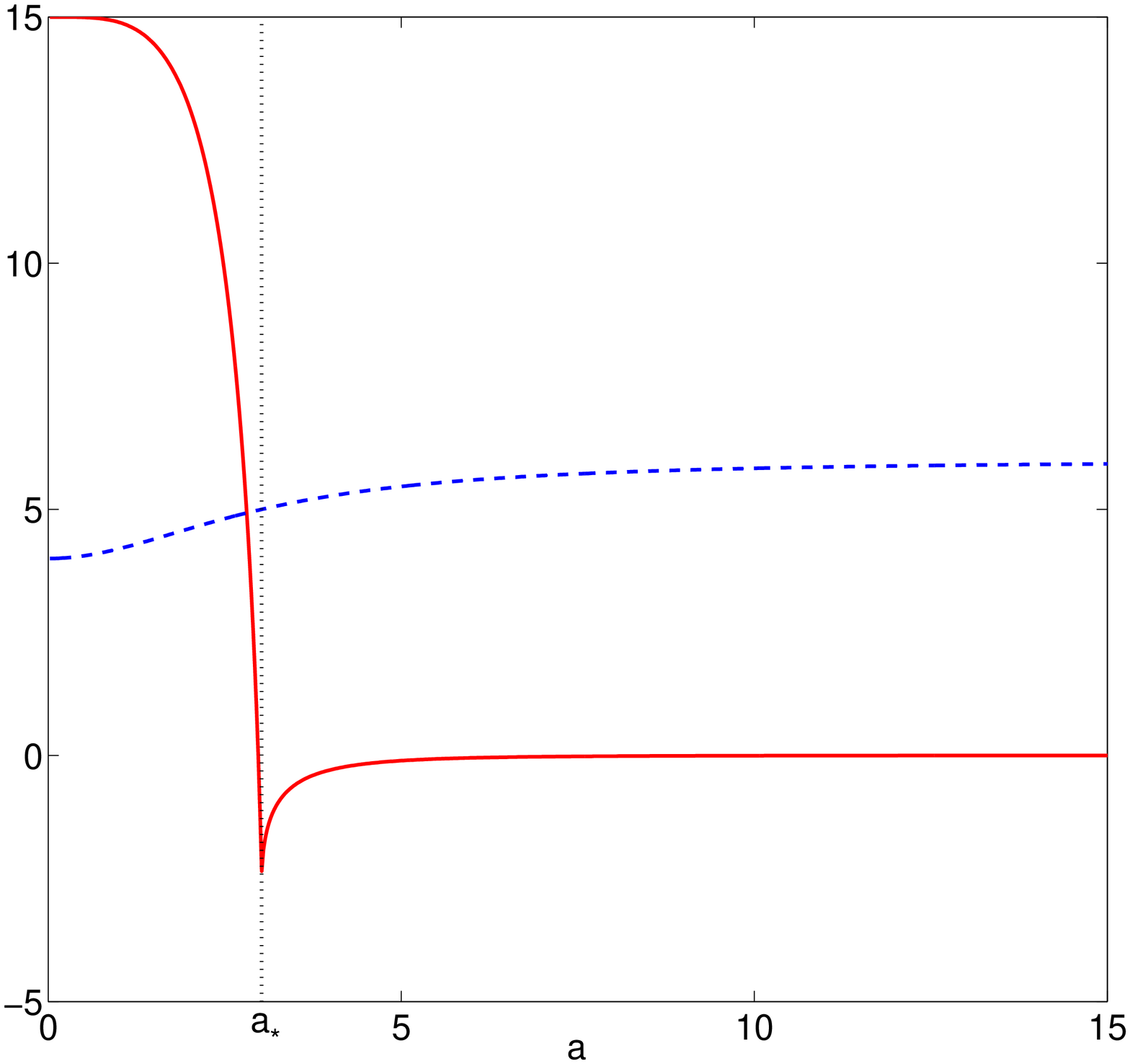}
\includegraphics[width=7.2cm]{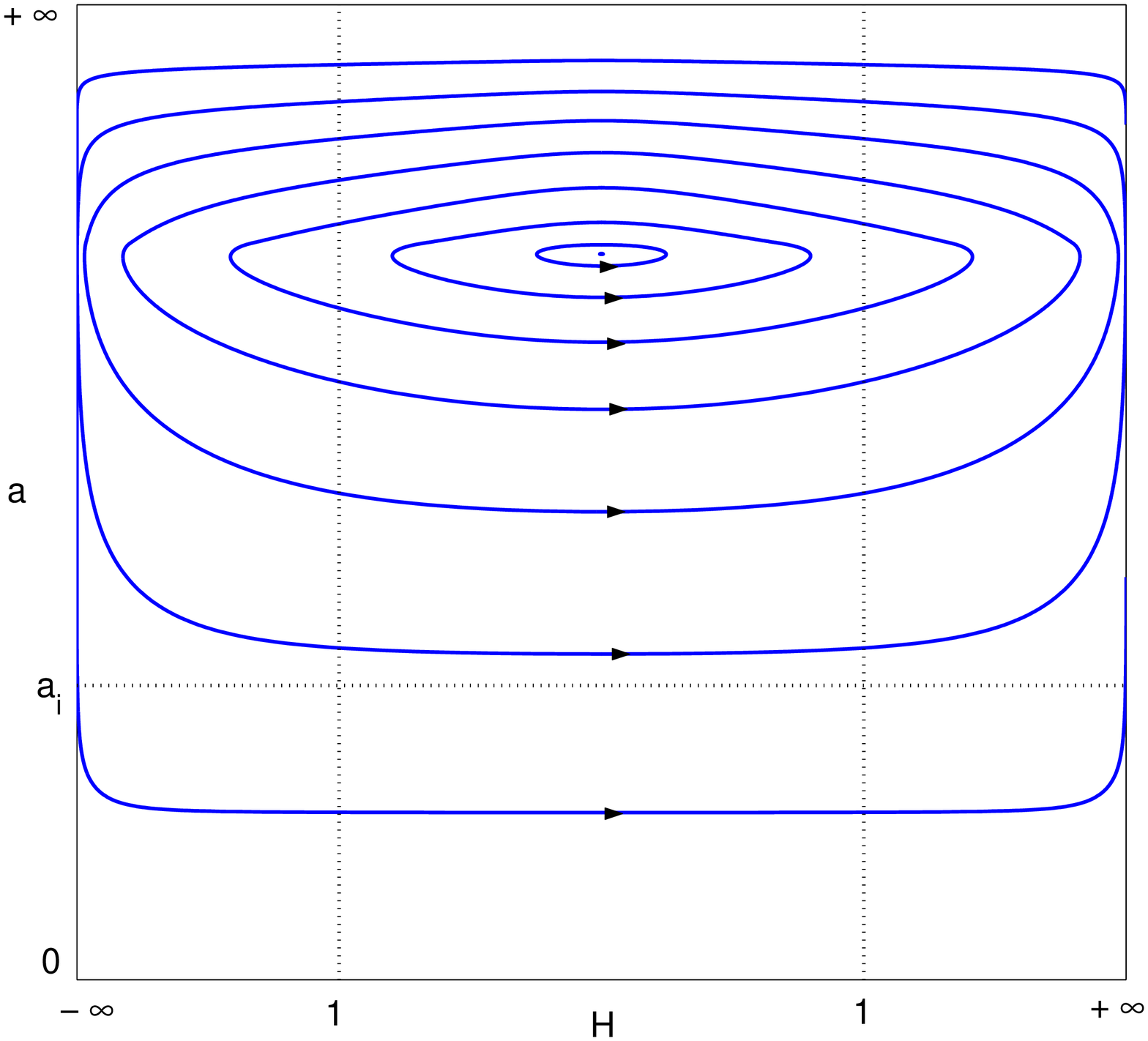}
\caption[] {\label{fig4.eps}
The top panel shows the plots of functions $A$ (solid line) and
$B$ (dashed line) against
$a$  for $V=-0.013$. The vertical dotted line 
denotes the position of $a=a_*$. 
The bottom panel is the corresponding phase portrait demonstrating the
centre equilibrium point that occurs in this case.
The plot is compactified as in Fig. \ref{fig2.eps}.
The dotted lines in the bottom panel
are as described in Fig. \ref{fig2.eps}.
All axes are in
Planck units and $j=100$. }

\end{figure}

$V<0$:
Negative potentials are known to arise in
string/M-theory compactifications (e.g., \cite{negative})
and are of interest
in connection with the ekpyriotic/cyclic models considered
recently \cite{cyclic}.
In such cases the qualitative change in the form of $B$ results
in only one equilibrium point, which occurs inside
the semi-classical region
and hence is a centre (see the bottom panel of Fig.~\ref{fig4.eps}).
Furthermore, the centre equilibrium point will
persist for {\em any} negative value of $V$.

To summarise, the important consequence of
considering LQC effects in this context is that they admit
two possible static solutions,
rather than the single solution in the case of
GR. The first static solution corresponds to a saddle point
(as in GR) and is referred to as the
Einstein Static (ES) solution. The second static solution is
a direct consequence of LQC effects. It is a
centre and we shall refer to it as the loop static (LS) solution.
We have shown analytically that the LS solution arises for a
wide range of values for the potential given by $V \in (-V_{lb}, V^*)$,
where the lower bound $-V_{lb}$ is determined by
the need to satisfy the Planck bounds.
The importance of the LS solution is that,
in contrast to the ES solution present in GR,
slight perturbations do not
result in an exponential divergence from the static universe, but
lead instead to oscillations about it.
Moreover, since the LS solution always lies
within the semi-classical region $a<a_*$, it is ideally
suited to act as the past asymptotic initial state for
an emerging universe. Indeed, it is
the only equilibrium point in models with a vanishing or negative
potential.

In the following Section, we
employ these novel features of LQC dynamics within the
context of the emergent inflationary universe.

\section{Emergent Inflationary Universe in LQC}

\subsection{The Dynamics of Emergence}

The phase plane analysis
of the previous Section determined the qualitative LQC
dynamics for the case of a constant potential.
However, any realistic inflationary model clearly requires the potential
to vary as the scalar field evolves. Nevertheless,
the constant potential dynamics (see Figs. \ref{fig2.eps}--\ref{fig4.eps})
provides a good approximation
to more general dynamics if variations in the potential
are negligible over a few oscillations. This implies that
cosmic dynamics with a variable potential may be studied
by treating the potential as a sequence of separate, constant potentials.
Here, with the emergent inflationary models in mind (see below),
we shall consider a general class of potentials
that asymptote to a constant $V_{-\infty} <V_*$ as
$\phi \rightarrow -\infty$ and rise monotonically once the value of
the scalar field exceeds a certain value.

As we saw in the previous Section, for any constant potential $V<V_*$,
there exists a region of parameter space
where the universe undergoes non--singular oscillations
about the point LS. For $V>V_*$, the equilibrium points LS and
ES merge and the phase plane then resembles that of classical GR:
a collapsing universe bounces and asymptotically evolves into a
de Sitter (exponential) expansion in the infinite future.
This leads us to propose the following picture for the origin of the
universe. The universe is initially at, or in the neighbourhood of, the
static point LS, with the field located on the plateau region of the
potential with a positive kinetic energy, $\dot{\phi}_{\rm init} >0$.
The universe undergoes a series of
non--singular oscillations in a (possibly) past--eternal phase
with the field evolving monotonically along the potential. As the
magnitude of the potential increases, the cycles are eventually broken
by the emergence of the universe into an inflationary epoch.

We proceed to discus the dynamics of this scenario in more detail.
The universe will exhibit cyclic behaviour around the LS point
for a very wide range of initial values $\dot{\phi}_{\rm init}$
when $V_{-\infty} <0$. If $0 < V_{-\infty}
< V_*$, the range becomes more limited as $V_{-\infty}$ is increased.
An important feature to note is that in all cases
the field's kinetic energy
never vanishes during a given cycle.
In the case of a positive potential,
the unphysical region of phase space where the field violates
the null energy condition lies outside the cyclic region
that is bounded by the separatrix. Indeed, for a constant potential,
the evolution of the scalar field is determined by integrating
Eq. (\ref{KGquantum}):
\begin{equation}
\label{intscalar}
\dot{\phi}= \dot{\phi}_{\rm init} \left( \frac{a_{\rm init}}{a}
\right)^3 \frac{D}{D_{\rm init}} \,.
\end{equation}
Thus, the field will vary monotonically along the potential
and eventually reach the region where the potential begins to
rise.

Increasing the magnitude of the potential over a series of
cycles has the effect of moving the location of the
LS point to progressively higher values of the scale
factor, although the shift is moderate. In the case of a positive potential,
for example, a necessary condition for the existence of LS is that
$4< A (a_{\rm eq}) <-2.5$,
where the function $A$ is defined in Eq. (\ref{AB}), and this corresponds to
the range $0.94 a_*< a < a_*$.
Eq. (\ref{KEfix}) then implies that the field's kinetic energy
does not alter significantly,
since the universe remains in the vicinity of LS.
On the other hand, the saddle point ES
occurs at progressively smaller values of the
scale factor as the magnitude of the potential increases.
As discussed in Section 2.2, this equilibrium point
occurs in the range $a_* \le a \le \infty$, where the limits
are approached as $V \rightarrow V_*$ and $V \rightarrow 0$, respectively.

The overall effect of increasing the potential, therefore, is to distort
the separatrix in the phase plane, making it narrower
in the vertical direction but introducing little change to the position
of the LS point. This implies that the dynamics varies only slightly from
cycle to cycle for orbits that are close to the LS point.
If the magnitude of the potential
continues to increase as the field evolves, however,
a cycle is eventually reached where the trajectory  that represents
the universe's evolution now lies {\em outside} the finite region
bounded by the separatrix and this effectively
breaks the oscillatory cycles. From a physical point of view,
the magnitude of the field's potential energy,
relative to its kinetic energy, is now sufficiently
large that a recollapse of the universe is prevented, i.e.,
the strong energy condition of GR is violated, thereby leading to accelerated
expansion. The field decelerates as it moves further up the potential,
subsequently reaching a point of maximal displacement and then
rolling back down. If the potential has a suitable form in this region,
slow--roll inflation will occur.

\subsection{The Energy Scale of Inflation}

An important question to address is the energy scale at the
onset of inflation, $V_{\rm inf}$.
In general, inflation begins (in the classical regime)
when the strong energy condition is violated:
\begin{equation}
\label{inflationcondition}
V(\phi_{\rm inf})
\approx \dot{\phi}^2_{\rm inf}\, ,
\end{equation}
and, moreover, the structure of the phase space indicates
that the potential energy of the
field remains dynamically negligible until the onset of
inflation \footnote{For a quadratic
potential, numerical integration of the field
equations indicates that this estimate yields a very good measure of
the energy scale at the onset of inflation when the universe undergoes
a large number of pre--inflationary oscillations \cite{mntl}.}.
In this case, the evolution of the scalar field prior to
inflation is determined by Eq. (\ref{intscalar}).

The inflationary energy scale may then
be estimated by considering the penultimate cycle before
the onset of inflation. The turnaround in the
expansion occurs when the field's energy density satisfies
$\rho =3/(8\pi \lpl^2  a^2)$. Since the energy
density of the field will not have changed significantly at the equivalent
stage of the following cycle, the
scale factor at the onset of inflation is determined approximately
by $a_{\rm inf} \approx (4\pi \lpl^2 V_{\rm inf})^{-\frac{1}{2}}$.
Substituting this condition into
Eq. (\ref{intscalar}) then yields an estimate for the magnitude of the
potential in terms of initial conditions:
\begin{equation}
\label{potapprox}
V_{\rm inf} \approx \frac{1}{\left( 4\pi \lpl^2 \right)^{\frac{3}{2}}}
\frac{D_{\rm init}}{a_{\rm init}^3 \dot{\phi}_{\rm init}} \,.
\end{equation}

For a universe located near to the equilibrium point
LS, the scale factor is given by $a_{\rm init} = fa_*$, where $0.94 \le
f \le 1$ and it may be further verified that 
$D_{\rm init}  = {\cal{O}} (1)$
in this range. The field's initial
kinetic energy is then determined by Eq. (\ref{KEfix}):
$\dot{\phi}^2_{\rm init} \approx 3/(4\pi \lpl^2 a^2_{\rm eq})$.
It follows from Eq. (\ref{potapprox}),
therefore, that a universe `emerging' from
the semi--classical LQC phase near to the LS
point will begin to inflate when
\begin{equation}
\label{staticstart}
V_{\rm inf} \approx \frac{1}{2j f^2} \mpl^4 \,.
\end{equation}
As expected, this is in good agreement with the
necessary condition (\ref{potentialmerge}) for the coalescence of the
equilibrium points LS and ES, since
the scale factor can not evolve until $V>V_*$ if
the universe is located on or near the LS point.

A precise measure on the set of initial data is presently
unknown, and we should therefore consider other possible initial
conditions. Having investigated the regime $a_{\rm init} 
\approx a_*$, a second 
possibility is to consider initial
conditions where $a_i \approx a_{\rm init} \ll a_*$. In this regime,
the quantum correction factor (\ref{defD}) asymptotes to a power--law,
$D \approx (12/7)^6 (a/a_*)^{15}$, and
Eq. (\ref{potapprox}) is then equivalent to the condition
\begin{equation}
\label{keinitial}
\dot{\phi}_{\rm init} \approx 20 \frac{\beta^{12}}{j^{\frac{3}{2}}}
\left( \frac{V_{\rm inf}}{m_{\rm Pl}^4} \right)^{-1} m_{\rm Pl}^2 \,,
\end{equation}
where we have defined the ratio $\beta \equiv a_{\rm init}/a_*$.
This ratio may be constrained by imposing
two necessary conditions for the semi--classical framework
to be valid. Firstly, the initial conditions should be set in the
regime where spacetime is approximated by a smooth manifold,
$a_i / a_{\rm init} < 1$, and this leads to the {\em lower}
limit:
\begin{equation}
\label{lowerbeta}
\beta > \sqrt{\frac{3}{j}} \,.
\end{equation}
Secondly, the scalar field's kinetic energy must not exceed the
Planck scale at the onset of the classical regime. Since
the anti--frictional effects in the modified field equation
(\ref{KGquantum}) accelerate the field when $a < a_*$, the
field's kinetic energy must initially be sub-Planckian,
$|\dot{\phi}_{\rm init} |/m_{\rm Pl}^2 \le 1$.
Eq. (\ref{keinitial}) then leads to an {\em upper} limit on $\beta$:
\begin{equation}
\label{upperbeta}
\beta <  \left( \frac{V_{\rm inf}}{m_{\rm Pl}^4}
\right)^{\frac{1}{12}} j^{\frac{1}{8}} \, ,
\end{equation}
and combining the constraints (\ref{lowerbeta}) and
(\ref{upperbeta}) results in a {\em necessary}
condition for the onset of inflation when $a_{\rm init} \ll a_*$:
\begin{equation}
\label{Vlimit}
\frac{V_{\rm inf}}{m_{\rm Pl}^4} > \left( \frac{2}{j}
\right)^{\frac{15}{2}} .
\end{equation}
In other words, for a given value of the
parameter $j$, inflation must begin above the
scale $(2/j)^{\frac{15}{2}} \mpl^4$.

Conditions (\ref{staticstart}) and (\ref{Vlimit}) both imply that the
inflationary energy scale is {\em higher}
for {\em lower} values of the parameter $j$.
Indeed, it is comparable to the Planck scale
for $j \le {\cal{O}} (10)$ and this generic behaviour is only weakly
dependent on the initial conditions.
This has implications for the asymptotic form of the
potential as the field reaches progressively higher values.
Unless the parameter $j$ is sufficiently large,
it is unlikely that the oscillatory dynamics will end if the
potential asymptotes to a constant value, or reaches a local
maximum, that is significantly below the Planck scale.
In this sense, therefore, the scenario outlined above
favours potentials that increase monotonically once the value of the scalar
field has exceeded some critical value.

In the following Section,
we consider a concrete example of a potential that exhibits
the appropriate asymptotic behaviour.

\begin{figure}[!t]
\includegraphics[width=7.2cm]{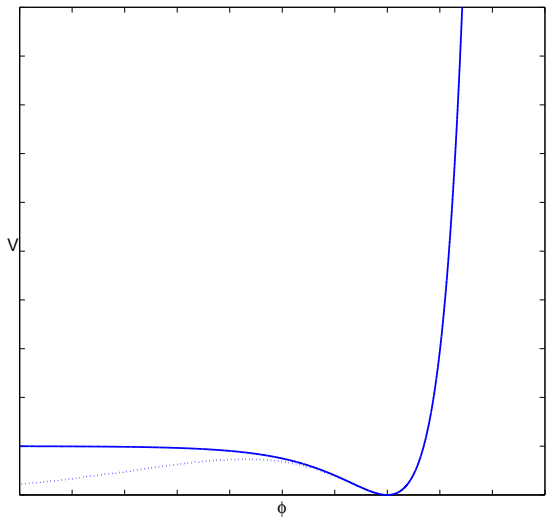}
\caption[] {\label{potential}
The figure depicts two possible forms of emergent potentials
that allow for conventional re-heating.  The solid line is the
form of the potential motivated by the inclusion of a $R^2$ term
in the Einstein--Hilbert action, and the dotted line the form motivated
by the inclusion of higher--order terms.
}
\end{figure}

\section{A Specific Model of an Emerging Universe}

From a dynamical point of view, the emergent
universe scenario can be realised within the context of
semi--classical LQC if the potential satisfies a number of
rather weak constraints. Asymptotically,
it should have a horizontal branch as $\phi \rightarrow
-\infty$ such that $dV/d \phi  \rightarrow 0$ and
increase monotonically in the region $\phi > \phi_{\rm grow}$,
where without loss of generality we may choose $\phi_{\rm grow} =0$.
The reheating of the universe imposes a further constraint that
there should be a global minimum in the potential at $V_{\rm min} =0$
if reheating is to proceed through coherent oscillations of the inflaton.
Since inflation will end after the field has rolled back down the potential,
this should occur for $\phi \le \phi_{\rm grow}$. Finally,
the region of the potential
that drove the last $60$ e--foldings
of inflationary expansion is then constrained by cosmological
observations, as in the standard scenario.

Examples of potentials that exhibit these generic properties are 
illustrated in Fig. \ref{potential}. It is interesting that
potentials of this form have been considered previously
in a number of different settings, including cases where
higher--order curvature invariants of the form
\be
L_{N}  =  \frac{1}{16 \pi \lpl^2} \sum_{i=1}^{N}
\epsilon_i\,R^{i} 
\ee
are introduced into the Einstein--Hilbert action, where
$R$ is the Ricci scalar, $\epsilon_i$ are coupling constants and $\epsilon_1=1$.
Such corrections are required when attempting to renormalize
theories of quantum gravity \cite{anttom86} and
also arise in low-energy limits of superstring theories \cite{canetal85}.
In general, such theories are conformally equivalent
to Einstein gravity sourced by a minimally coupled, self--interacting
scalar field. For example, potentials with a nonzero
asymptote at $\phi \rightarrow -\infty$
(as shown by the solid line in Fig. \ref{potential})
can be obtained from theories that include a quadratic term
in the action, whereas those with a
zero asymptotic value and a local maximum (shown by a dotted line)
arise when cubic and higher--order terms in the Ricci
scalar are introduced \cite{maeda89,bc91}.
In general, all these potentials possess
a global minimum at $V=0$.
Finally, potentials
of this form have also been considered within the context of the classical
emergent universe \cite{Ellis-Murugun-Tsagas04}.

Motivated by the above discussion, we consider, as an example, the potential
\be
\label{pot}
V= \alpha \left[ ( \exp ( \beta \phi / \sqrt{3} )-1   \right]^2 \,,
\ee
where $\alpha$ and  $\beta$ are constants.
This potential is qualitatively similar to that illustrated
by the solid line in Fig.~\ref{potential}.
The parameters in the potential are constrained primarily
by the Cosmic Microwave Background (CMB) anisotropy
power spectrum. Assuming inflation proceeds in the region $\phi  \gg 0$,
the parameter $\beta$ determines the spectral index
of the scalar perturbation spectrum together with the
relative amplitude of the gravitational wave perturbations.
The parameter $\alpha$ is then constrained by the COBE
normalization of the power spectrum on large--scales \cite{COBEnorm}.

\begin{figure}
\includegraphics[width=7.2cm]{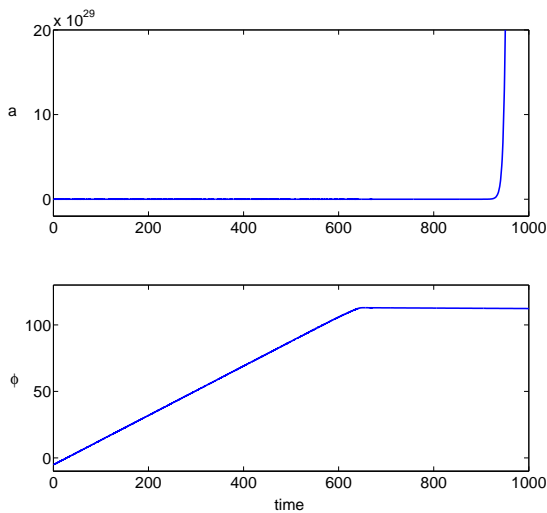}%
\caption[] {\label{fig5.eps}
Time evolution of the scale factor (top panel) and scalar
field (bottom panel) with the field
initially on the asymptotically flat region of the potential
(\ref{pot}) with $\alpha = 10^{-12}$ and $\beta =0.1$. The field 
increases in value from initial conditions close to the LS
(centre) solution defined by Eq. (\ref{eq-points}).
}
\end{figure}

\begin{figure}
\includegraphics[width=7.2cm]{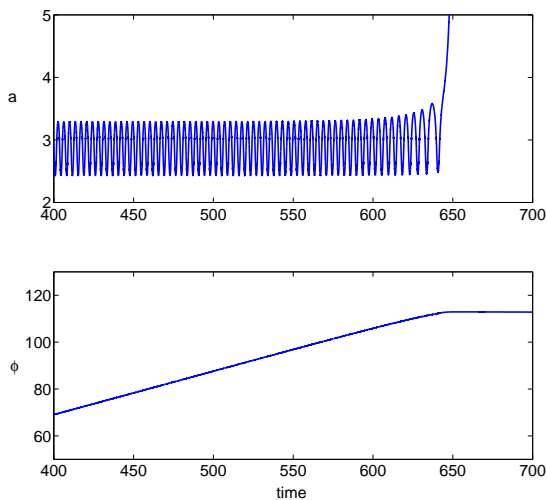}%
\caption[] {\label{fig6.eps}
Plots illustrating the magnification of 
Fig.~\ref{fig5.eps} around the region
where emergence commences,
the oscillations cease and inflation begins.
}
\end{figure}

We chose $\alpha=10^{-12}$
and $\beta = 0.1$ as representing typical
values satisfying the constraints imposed recently by
the WMAP satellite \cite{wmap} and numerically integrated the
field equations (\ref{Rayquantum})--(\ref{Friedquantum}) for
a universe starting from an initial state close to the LS
static state.
The results of the integration are
illustrated in Figs. \ref{fig5.eps} and \ref{fig6.eps}.
The field starts in the asymptotically flat region of the
potential and gradually increases in value, as the scale
factor oscillates about the LS point. The field moves past the minimum
and climbs up the potential. The scale factor continues
to oscillate until the field reaches the point where it slows
down significantly, thereby bringing the oscillations to an
end and initiating the inflationary expansion. The behaviour at this stage
is qualitatively similar to that discussed
previously for a quadratic potential \cite{lmnt}.

\section{Emerging Quintessential Inflation}

One drawback of reheating through inflaton decay in the emergent
scenario is that the coupling of the scalar field to the standard
model degrees of freedom must be strongly suppressed
if the field is to survive for a (possibly) infinite time
as it emerges from the oscillatory semi--classical phase. It is
more natural, therefore, to invoke a `sterile' inflaton
that is not coupled directly to standard model fields, and
where reheating proceeds through an alternative mechanism
such as gravitational particle production \cite{gpp}. In this case,
the potential need not exhibit a minimum and the field
could continue to roll back down the potential towards $\phi \rightarrow
-\infty$ at late times.

It is notable that the general requirements for a sterile
inflaton with a potential exhibiting a decaying tail as $\phi \rightarrow
-\infty$ are {\em precisely} those features that are characteristic
of the quintessential inflationary scenario, where the
field that drove early universe inflation is also
identified as the source of dark energy today \cite{PV}. 
In the present context,
this suggests that the scalar field could play a three--fold
role in the history of the universe, acting as the mechanism that
enables the universe to emerge into the classical domain, and then
subsequently driving both the early-- and late--time accelerated expansion.

The specific constraints that must be
satisfied by the potential in standard quintessential
inflation have been considered in detail in Ref. \cite{quin}.
In particular, one of the simplest asymptotic forms for
the low--energy tail that simultaneously leads to tracking behaviour and
satisfies primordial nucleosynthesis bounds
is given by $V \propto (m/\phi)^k \exp (\lambda \phi /m_{\rm Pl})$,
where $ m, k$ and  $\lambda$ are constants. Cyclic behaviour in LQC
will arise for such a potential.

\begin{figure}
\includegraphics[width=7.2cm]{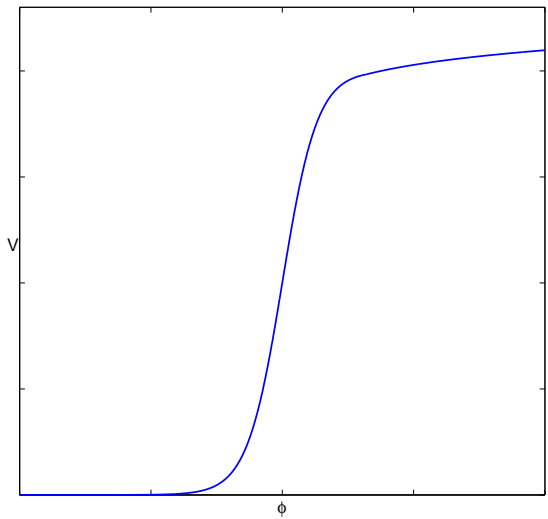}%
\caption[] {\label{potential2}
Plots illustrating the generic form of the potential that leads to early-- and
late--time accelerating phases. The potential exhibits a decaying tail 
as $\phi \rightarrow -\infty$. As the field moves up 
this tail and increases in value, the universe can oscillate about the LS 
point. The inflationary regime rises (possibly) towards the Planck scale 
for large values of $\phi$. As the field turns round, it can drive a phase of 
inflation and, if the potential exhibits a sufficiently 
steep middle section around $\phi \approx 0$, reheating may 
proceed through gravitational particle
production. Consequently, the field may survive until the present epoch,
where it can act as the source of dark energy by slowly rolling
along the tail towards $\phi \rightarrow -\infty$.
}
\end{figure}

A further requirement is that the potential must be sufficiently
steep immediately after the end of inflation if the field's energy
density is to redshift more rapidly than the sub--dominant
radiation component. This requires a second point of inflexion in the
potential, as illustrated qualitatively in Fig. \ref{potential2}. Beyond this
region, the potential must continue to rise in order for the
oscillatory dynamics to come to end and, as discussed above,
this is expected to occur near to the Planck scale. From
a dynamical point of view, there are no further constraints
on how rapidly the potential energy need increase in this region.
The only remaining
consideration is that a phase of successful slow--roll inflation
should arise as the field rolls back down the potential.
Given the ease with which the inflaton is
able to move up the potential due to LQC effects,
we anticipate that any potential satisfying the existing constraints
for successful quintessential inflation will also lead to
a successful emergence of the classical universe.

\section{Discussion}

In this paper, we have investigated the occurrence of
static solutions in loop quantum cosmology settings sourced
by a scalar field with a constant potential.
We have shown that there are in principle
two such solutions, depending upon the sign and magnitude of the
potential. The point ES is always in the classical domain
and corresponds to the unstable saddle point that is also present in GR.
The important characteristic of this
solution is that any perturbations, no matter how small,
necessarily force the
universe to deviate exponentially from the static state.
The second solution LS is always
in the semi-classical domain and corresponds to a centre.
This is a solution made possible by LQC effects. We have shown that it exists
for a wide range of values of the potential,
including positive, zero and negative values.
Its importance lies in the fact that
it allows a universe that is slightly displaced from the static state
to oscillate in the neighbourhood of the
static solution for an arbitrarily long time.

We have exploited these characteristics to develop a working scenario
of the emerging inflationary universe, in which a past--eternal,
cyclic cosmology eventually enters a phase where the symmetry of
the oscillations is broken by the scalar
field potential, thereby leading in principle to a phase of successful
slow--roll inflation. The mechanism that enables the universe
to emerge depends very weakly on the form of the potential, and
requires only that it asymptotes to a constant
value at $\phi \rightarrow -\infty$ and grows in magnitude at larger $\phi$
in order to break the cycles. The asymptotic value of the potential
can be either positive, zero or negative.

The above emergent scenario has a further advantage
that the initial state of the universe is set in the
more natural semi-classical regime, rather than the classical
arena of GR. Nevertheless,
an important question that arises is the likelihood of
these initial conditions within a more general framework.
In particular, there is the issue of why the scalar field
should initially be located in the asymptotic low-energy 
region of the potential. 
Although a detailed study of such questions is beyond the
scope of the present paper, it has been argued \cite{martin_kevin_1}
that for the case of a constant potential, the wavefunction
in LQC most closely resembles the Hartle--Hawking no--boundary
wavefunction \cite{hh}. 
More specifically, the difference equation in LQC
requires the wavefunction to tend to zero near to the classical singularity
\cite{bojoic},
and in this sense resembles DeWitt's initial condition \cite{dewitt}. 
However,
within the context of solutions to the Wheeler--DeWitt equation,
requiring the wavefunction to be bounded as $a \rightarrow 0$
selects the exponentially increasing WKB mode \cite{martin_kevin_1} and this
corresponds to the no--boundary wavefunction.

This is of interest since the square of the wavefunction
in quantum cosmology is interpreted as
the probability distribution for initial values of the scalar
field in an ensemble of universes. For the no--boundary
boundary condition, the probability,
$P \propto \exp( 3/[8\pi l_P^2V(\phi )]) $, is peaked at $V(\phi ) =0$,
thereby suggesting that the no--boundary proposal disfavours
inflation. In the present context, however, this
implies that the most natural initial condition
for the field is to be located either at the minimum of the
potential, or in the case where the potential has no
minimum, at $\phi =- \infty$.
It would be interesting to explore this possibility further.
In particular, previous analyses have so far neglected
kinetic terms in the matter Hamiltonian and these may be important
in the emerging universe scenario.
There is also the related question of whether the field moves from
left to right initially.

In general, we have found that the onset of slow--roll inflation
will occur at a relatively high energy scale,
unless the quantization
parameter $j$ is extremely large. This indicates that a large amount
of slow--roll inflation should arise, at least for a wide class of
smoothly varying potentials, and it is expected, therefore, that
the density of the present--day universe should be exponentially
close to the critical density, $\Omega_0 =1+\epsilon$, where
$\epsilon \ll 1$. In principle, therefore, the emergent
scenario we have proposed could be ruled out if a significant
detection of spatial curvature is ultimately reported by future
cosmological observations.

It is also worth remarking that if the field is pushed
sufficiently far up the potential, perhaps as high as
the Planck scale, the inflating universe may
enter a phase of eternal self--reproduction, where quantum
fluctuations in the inflaton become more important than its
classical dynamics \cite{eternal}.

Finally, there is an interesting symmetry
in the emerging quintessential scenario between the
initial and final states of the universe. Although
the size of the universe differs by many orders of
magnitude, the field evolves along the tail of the potential at
both early and late times, $\phi (t \rightarrow -\infty )
= \phi (t \rightarrow +\infty )$, but with its kinetic energy
having changed sign.
This implies that a reconstruction of the dark energy equation of state
at the present epoch could yield direct
observational insight into the nature of the {\em pre--inflationary}
potential in this scenario.

\section*{Acknowledgments}  
DJM is supported by PPARC.

\end{document}